\newcommand{\eps}{\epsilon}

\newcommand{\sx}{\sigma_x}
\newcommand{\sy}{\sigma_y}
\newcommand{\sz}{\sigma_z}

\newcommand{\bra}[1]{\left| #1 \right \rangle}

\newcommand{\U}{\uparrow}
\newcommand{\D}{\downarrow}
\newcommand{\tikzcircle}[2][red,fill=red]{\tikz[baseline=-0.5ex]\draw[#1,radius=#2] (0,0) circle ;}%

\documentclass[%
reprint,amsmath,amssymb,aps,prl,superscriptaddress,
]{revtex4-1}

\usepackage{graphicx}
\usepackage{dcolumn}
\usepackage{bm}
\usepackage{epstopdf}
\usepackage{float}
\usepackage{tikz}
\usepackage{MnSymbol}
\usepackage{wasysym}

\begin{document}

\title{Coherent spin state transfer via Heisenberg exchange}

\author{Yadav P. Kandel}
\thanks{These authors contributed equally.}

\author{Haifeng Qiao}
\thanks{These authors contributed equally.}

\affiliation{Department of Physics and Astronomy, University of Rochester, Rochester, NY, 14627 USA}

\author{Saeed Fallahi}
\affiliation{Department of Physics and Astronomy, Purdue University, West Lafayette, IN, 47907 USA}
\affiliation{Birck Nanotechnology Center, Purdue University, West Lafayette, IN, 47907 USA}

\author{Geoffrey C. Gardner}
\affiliation{Birck Nanotechnology Center, Purdue University, West Lafayette, IN, 47907 USA}
\affiliation{School of Materials Engineering, Purdue University, West Lafayette, IN, 47907 USA}

\author{Michael J. Manfra}
\affiliation{Department of Physics and Astronomy, Purdue University, West Lafayette, IN, 47907 USA}
\affiliation{Birck Nanotechnology Center, Purdue University, West Lafayette, IN, 47907 USA}
\affiliation{School of Materials Engineering, Purdue University, West Lafayette, IN, 47907 USA}
\affiliation{School of Electrical and Computer Engineering, Purdue University, West Lafayette, IN, 47907 USA}

\author{John M. Nichol}
\email{jnich10@ur.rochester.edu}

\affiliation{Department of Physics and Astronomy, University of Rochester, Rochester, NY, 14627 USA}

\begin{abstract}
Quantum information science has the potential to revolutionize modern technology by providing resource-efficient approaches to computing~\cite{Ekert1996}, communication~\cite{Kimble2008}, and sensing~\cite{Degen2017}. Although the physical qubits in a realistic quantum device will inevitably suffer errors, quantum error correction creates a path to fault-tolerant quantum information processing~\cite{Knill2005}. Quantum error correction, however, requires that individual qubits can interact with many other qubits in the processor. Engineering this high connectivity can pose a challenge for platforms like electron spin qubits~\cite{Ladd2010} that naturally favor linear arrays. Here, we present an experimental demonstration of the transmission of electron spin states via Heisenberg exchange in an array of spin qubits. We transfer both single-spin and entangled states back and forth in a quadruple quantum-dot array without moving any electrons. Because it is scalable to large numbers of qubits, state transfer through Heisenberg exchange will be especially useful for multi-qubit gates and error-correction in spin-based quantum computers. 
\end{abstract}



\maketitle

Spin qubits based on electrons in quantum dots are a leading platform for quantum information processing, because the quantum phase coherence of individual electron spins can persist for extremely long times ~\cite{Loss1998,Kane1998}. Single-qubit gate fidelities now exceed 99.9$\%$~\cite{Yoneda2018,Chan2018,Muhonen2015} and two-qubit gate fidelities exceed 98$\%$~\cite{Huang2019}. As spin-based quantum processors scale up, one- and two-dimensional arrays of electrons in quantum dots have emerged as key components of future spin-based quantum information processors~\cite{Zajac2016,Mortemousque2018,Mukhopadhyay2018}. 

Electron spin qubits most naturally interact with each other via direct wavefunction overlap, which generates Heisenberg exchange coupling~\cite{Loss1998}. In large-scale arrays of spin qubits~\cite{Zajac2016,Volk2019}, however, maintaining sufficient connectivity for efficient and fault-tolerant quantum computing poses a challenge. To this end, long-distance coupling between spins is an active area of research. Exciting possibilities include coupling spins to superconducting microwave photons~\cite{Mi2018,Landig2018,Samkharadze2018}, shuttling electrons between quantum dots via tunneling~\cite{Mills2019,Fujita2017,Flentje2017,Nakajima2018,Baart2016,Greentree2004} or surface acoustic waves~\cite{Shilton1996,Bertrand2016}, and superexchange methods~\cite{Baart2016S,Malinowski2019}.  Theoretical proposals have also explored the use of repeated SWAP operations between qubits to achieve this goal~\cite{Aharonov1997,Gottesman2000,Fowler2004} and also the possibility of an exchange-based spin bus~\cite{Friesen2007}. 

In this work, we present an experimental demonstration of spin-state transfer via Heisenberg exchange coupling. We transmit the spin state of an electron back and forth across a quadruple quantum dot array, without ever moving any electrons. We also transfer one spin of an entangled pair to a distant electron and back. In contrast to previous work using electron tunneling~\cite{Mills2019,Fujita2017,Flentje2017,Nakajima2018,Baart2016,Greentree2004}, our approach relies entirely on coherent SWAP operations between spins and does not involve the motion of electrons. As a result, it is compatible with arbitrary single- and multi-qubit states. This scheme does not require separate entities such as microwave resonators, magnetic gradients, additional electrons, or empty quantum dots. State transfer via Heisenberg exchange is also scalable to large arrays of qubits, an essential requirement for quantum error correction. 

\begin{figure}
	\includegraphics{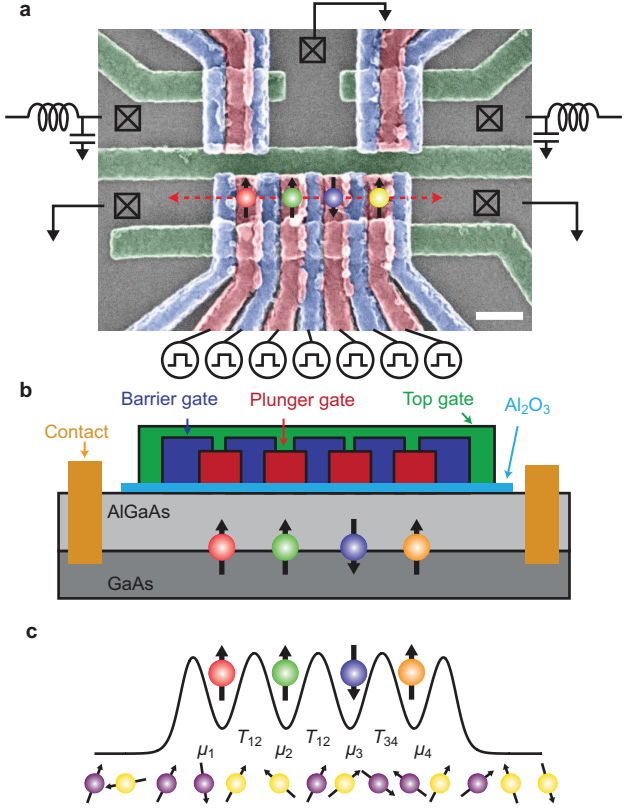}
	\caption{\label{apparatus} Experimental setup. (a) Scanning electron micrograph of the quadruple quantum dot.  The scale bar is 200 nm. All plunger gates (red) and inner barrier gates (blue) are connected to separate arbitrary waveform channels for independent pulsing. Ohmic contacts to the two-dimensional electron gas are indicated by an ``x" inside of a square. The two quantum dots above the middle green gate are charge sensors, and their ohmic contacts are configured for rf-reflectometry. A grounded top gate (not shown) covers the active area of the device. (b) Line cut through the device at the position indicated in panel (a) showing the locations of all electrons.  (c) Schematic potential landscape imposed by the confinement gates. In general, the plunger gates primarily control the chemical potential $\mu_i$ of each dot $i$, and the barrier gates primarily control the tunnel coupling between dots $i$ and $j$, $T_{ij}$. Ga and As nuclear spins (yellow and purple) contribute a random magnetic field at the site of each dot via the hyperfine interaction.}
\end{figure}

\begin{figure}
	\includegraphics{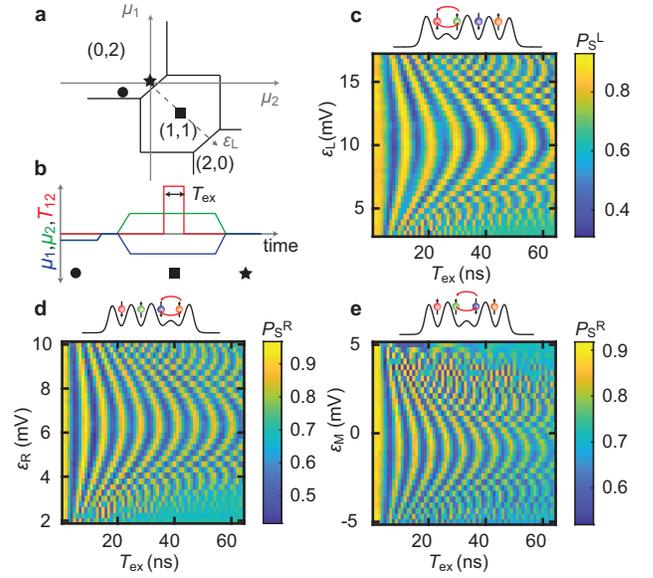}
	\caption{\label{exchange} Coherent exchange oscillations between all nearest-neighbor pairs of spins. (a) Schematic charge stability diagram for dots 1 and 2 showing the initialization (\tikzcircle[black, fill=black]{2pt}), readout ($\star$), and manipulation ($\blacksquare$) chemical potential configurations. The chemical potentials $\mu_1$ and $\mu_2$ and detuning $\eps_L$ are defined to be zero near the (0,2)-(1,1) transition (see Methods). (b) Pulse timing diagram for exchange measurements on dots 1-2. Typical initialization times are $2\mu$s, exchange pulse times $T_{ex}<100$ns, measurement times are $5\mu$s.  Overall pulse repetition periods are $<30 \mu$s. (c) Exchange oscillations between spins in dots 1 and 2.  (d) Exchange oscillations between spins in dots 3 and 4.  (e) Exchange oscillations between spins in dots 2 and 3. Oscillations were measured on the right side of the array. The visibility in this case is not as high as the other two panels, because we did not stabilize the magnetic gradient on the right side for this measurement. In (c)-(e), $P_S^{L(R)}$ indicates the singlet return probability for the left(right) side, and the initial states are shown at the top of each panel. To generate exchange coupling between dots $i$ and $j$, we pulse the barrier gate $T_{ij}$ for a time $T_{ex}$.}
\end{figure}

We use a quadruple quantum dot in a GaAs/AlGaAs heterostructure [Figs.~\ref{apparatus}(a)-(b)]. The device has an overlapping-gate architecture (see Methods), which enables precise control of the electronic confinement potential. For these experiments, each dot contains one electron. Using ``virtual gates''~\cite{Baart2016,Mills2019,Volk2019}, we independently control the chemical potentials of the quantum dots and the tunnel barriers between them [Fig.~\ref{apparatus}(c) and Figs.~\ref{exchange}(a)-(b)]. 

We initialize and measure the array by configuring it as a pair of singlet-triplet qubits (see Methods). Each singlet-triplet qubit occupies a pair of quantum dots. In the following, we refer to the left and right sides of the quadruple dot. We can initialize either side as $|\uparrow \uparrow\rangle$, $|\downarrow \uparrow\rangle$, or $|S\rangle=\frac{1}{\sqrt{2}} \left( |\uparrow \downarrow\rangle - |\downarrow\uparrow \rangle \right)$. The orientation of the spins in the $\bra{\D \U}$ state depends on the local magnetic gradient, which results from the hyperfine interaction between the electron and nuclear spins (see Methods). We measure both sides of the array via standard spin-to-charge conversion through Pauli spin blockade~\cite{Petta2005}. We readout in the $\{ |S\rangle,|T\rangle\}$ basis for each side of the array, where $|T\rangle$ is any one of the triplet states $\{| \uparrow \uparrow\rangle,\frac{1}{\sqrt{2}} \left( |\uparrow \downarrow\rangle + |\downarrow\uparrow \rangle \right), |\downarrow \downarrow \rangle\}$.  Adiabatic charge transfer of the electrons on each side into dots 1 and 4 maps $|\downarrow \uparrow\rangle \to |S\rangle$, and all other product states to triplets. Diabatic charge transfer from the outer dots preserves the spin states, and diabatic transfer into the outer dots during readout projects a joint spin state onto the $\{ |S\rangle,|T\rangle\}$ basis ~\cite{Petta2005}.

\begin{figure*}
	\includegraphics{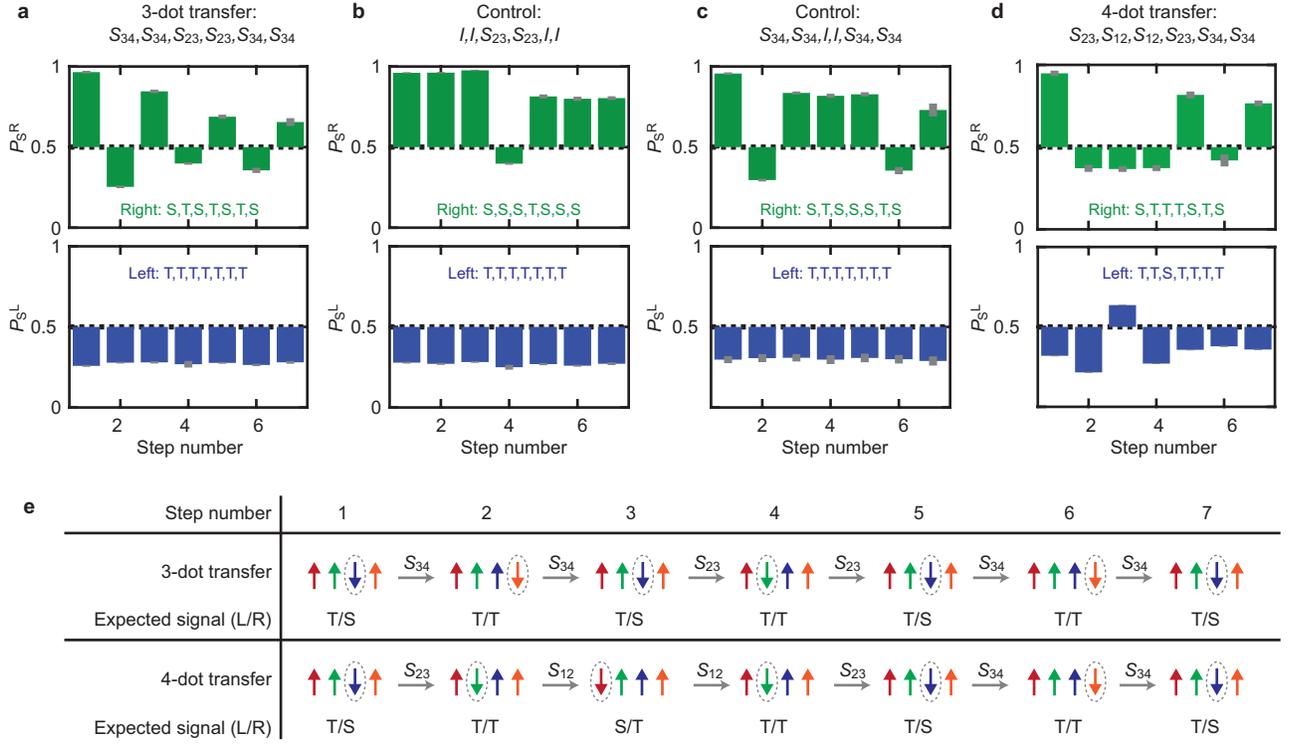}
	\caption{\label{teleport} Spin-state transfer via Heisenberg exchange. (a) Data for the three-dot state transfer described by the sequence $S_{34}, S_{34}, S_{23}, S_{23}, S_{34}, S_{34}$. (b) Three-dot state transfer control sequence with $I$ in place of $S_{34}$. (c) Three-dot control sequence with $I$ in place of $S_{23}$. (d) Four-dot state transfer described by the sequence $S_{23}, S_{12} , S_{12} , S_{23} , S_{34} , S_{34}$. In panels (a)-(d), the upper graph shows measurements on the right side, and the lower graph shows measurements on the left side. The insets gives the expected outcomes. (e) Trajectories of the down spin for the three- and four-dot state transfer sequences. In panels (a)-(d), gray bars are error bars, which represent the standard deviation of 64 repetitions of the average of 64 single-shot measurements of each pulse.}
\end{figure*}

We induce exchange coupling between two electrons by applying a voltage pulse to the barrier gate between them (see Methods) [Fig.~\ref{exchange}(b)]. The voltage pulse creates an overlap between the wavefunctions of neighboring electrons, and the spins evolve according to the Heisenberg exchange Hamiltonian $H_{int} = h \frac{J}{4} (\sx \otimes \sx+\sy \otimes \sy+\sz \otimes \sz)$. Here $J$ is the coupling strength, $\sigma_x$, $\sigma_y$, and $\sigma_z$ are Pauli matrices describing the spin components of each electron, and $h$ is Planck's constant. After the two-electron system evolves for a time $\frac{1}{2J}$, an initial state $|\psi\rangle $ evolves to $U|\psi \rangle$, where

\begin{align}
	U\propto
	\left(\begin{matrix}
		1 & 0 & 0 & 0\\
		0 & 0 & 1 & 0\\
		0 & 1 & 0 & 0\\
		0 & 0 & 0 & 1
	\end{matrix}\right).
\end{align}
$U$ is written in the basis $\{|\uparrow \uparrow\rangle, |\uparrow \downarrow\rangle,|\downarrow \uparrow\rangle, |\downarrow \downarrow\rangle\}$. $U$ describes a SWAP operation between the two spins. If the two are opposite, they swap back and forth as they evolve for a variable amount of time under the action of this Hamiltonian, generating exchange oscillations. As discussed further below, the addition of single-qubit terms to this Hamiltonian, especially magnetic field differences between spins, can lead to errors in the SWAP operation. Figures~\ref{exchange}(c)-(e) demonstrate coherent exchange oscillations between all nearest-neighbor pairs of spins in the array. 

\begin{figure*}
	\includegraphics{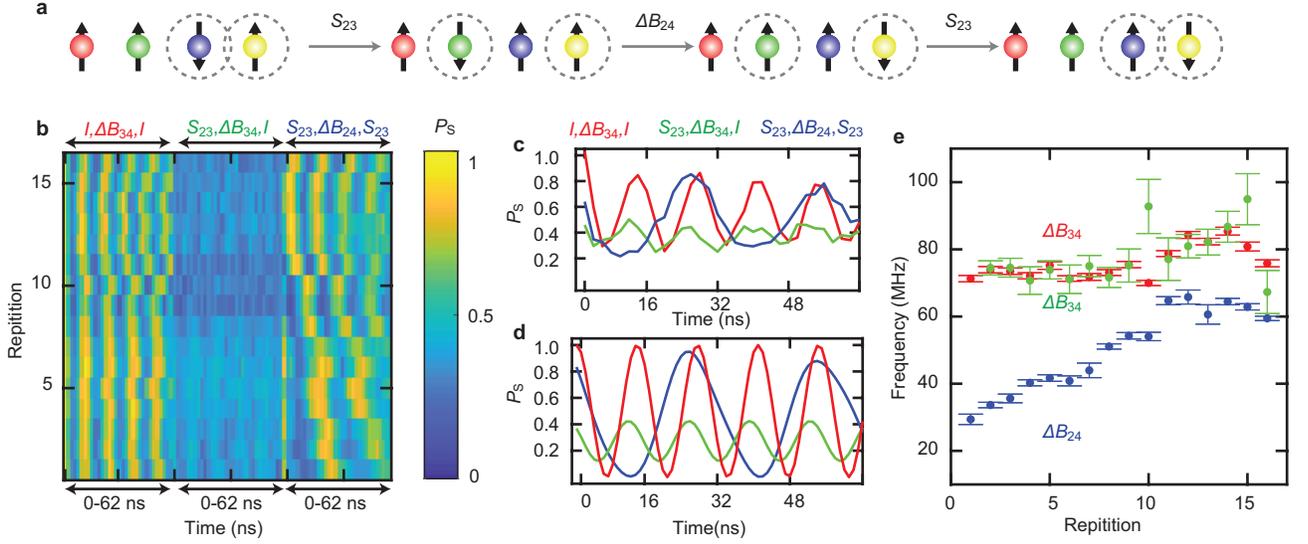}
	\centering
	\caption{\label{entangled} Transfer of entangled states via Heisenberg exchange. (a)  We prepare a singlet in dots 3 and 4 and then swap the states in dots 2 and 3. The separated singlet evolves in a magnetic field difference $\Delta B_{24}$. We then swap the state of dot 2 back into dot 3 and measure the right side. (b) Entangled-state transfer data. Three measurements are interleaved in time. The first, denoted $I,\Delta B, I$ is a control measurement keeping the singlet state in dots 3 and 4. The second is also a control measurement  $S_{23},\Delta B, I$, which swaps the singlet into dots 2 and 4 but omits the second $S_{23}$ before measurement. The third, denoted $S_{23},\Delta B, S_{23}$ contains both SWAP operations and demonstrates transfer of entangled states. For each line, we interleaved the averaging of these measurements, as described in Methods. (c) Data from repetition 2. Clear oscillations are evident in the $S_{23},\Delta B, S_{23}$ case, demonstrating transfer of entangled states. (d) Simulation corresponding to the measurements in (c). (e) Measured values of $\Delta B_{24}$ and $\Delta B_{34}$ vs. repetition number obtained by fitting the relevant section of data to a sinusoid. $\Delta B_{24}$ and $\Delta B_{34}$ show different trends in time, because they result from different nuclear spin configurations. Error bars are fit errors.}
\end{figure*}

To transfer the spin state of an electron, we initialize the array in the $|\uparrow \uparrow \downarrow \uparrow\rangle$ state, and we concatenate different SWAP operations between pairs of electrons. To swap the spins of the electrons in dots 3 and 4, for example, we apply a voltage pulse to barrier gate $T_{34}$, timed to give a $\pi$ pulse. We denote this operation as $S_{34}$. In general, we use $S_{ij}$ to mean a SWAP operation between spins $i$ and $j$. During each barrier pulse, we apply compensation pulses to the plunger gates such that the chemical potentials of the dots themselves remain fixed, and the electrons do not move. Typically, exchange pulses are $<10$ns in length and are usually 3$\pi$ pulses to ensure that exchange strengths are larger than magnetic gradients, as discussed below. 

We begin by transmitting the down spin originally associated with the electron in dot 3 through the following sequence of operations: $S_{34}, S_{34} , S_{23} , S_{23} , S_{34} , S_{34}$. Before the sequence begins and after each step, we measure both sides of the array (in the $\{ |S\rangle ,|T\rangle\}$ basis) to confirm the expected spin states [Fig.~\ref{teleport}(a)]. For this sequence, as shown in Fig.~\ref{teleport}(e), we expect the following measurement outcomes on the right side: $S,T,S,T,S,T,S$ and the following sequence on the left: $T,T,T,T,T,T,T$. The data in Fig.~\ref{teleport}(a) clearly show the expected outcomes, although the visibility of the measurements decreases with each successive step. We discuss the limiting factors in this state-transfer process below. We emphasize that the electrons themselves do not move during this process. It is only the spin-down state, which was originally associated with the electron in dot 3, that moves. The imperfect visibility of the prepared triplet states on the left side is due to thermal population of excited spin states, as discussed in the Methods section.

As a check, we have verified that eliminating certain pulses in the state transfer sequence also produces the expected results. Figure~\ref{teleport}(b) displays the outcome when we replace $S_{34}$ with the identity operation $I$, implemented as a wait with no barrier pulse. The data show the expected result. Likewise, we have checked that replacing $S_{23}$ with $I$ also gives the expected result in Fig.~\ref{teleport}(c).

We also demonstrate that we can transmit a spin back and forth across the full four-dot array. We apply the following swap sequence: $S_{23}, S_{12}, S_{12} , S_{23}, S_{34}, S_{34}$. For this sequence, we expect the following measurements on the right side: $S,T,T,T,S,T,S$ and the following on the left: $T,T,S,T,T,T,T$. The expected trend is clearly evident in the data [Fig.~\ref{teleport}(d)].

Having established the feasibility of transmitting single-spin eigenstates, we now demonstrate transmission of entangled states (Fig.~\ref{entangled}). By electronic exchange with the reservoirs and diabatic charge transfer, we prepare the array in the $\frac{1}{\sqrt{2}}|\uparrow\uparrow\rangle \otimes \left( |\uparrow \downarrow\rangle -|\downarrow \uparrow\rangle \right)$ state. Then, we apply an $S_{23}$ operation [Fig.~\ref{entangled}(a)]. This operation causes the singlet state, which was initially prepared in dots 3 and 4, to reside in dots 2 and 4. 

In general, a separated singlet state in dots $i$ and $j$ will evolve to the unpolarized triplet state $|T_0\rangle$ and back, if there exists a magnetic-field difference $\Delta B_{ij}$ between quantum dots $i$ and $j$~\cite{Petta2005}.  The singlet-triplet oscillation frequency is $g \mu_B \Delta B_{ij}/h$, where $g$ is the electron $g$ factor, $\mu_B$ is the Bohr magneton, and $h$ is Planck's constant. In our experiment, after the separated singlet state evolves for a variable period of time around the magnetic gradient, we apply an $S_{23}$ operation, bringing the singlet back into dots 3 and 4. We then measure the right side of the device in the usual $\{ |S\rangle, |T\rangle \}$ basis after diabatic charge transfer. Provided the $S_{23}$ operations preserve the entangled state, we expect to observe coherent singlet-triplet oscillations, corresponding to evolution around $\Delta B_{24}$. 

We observe clearly-visible singlet-triplet oscillations [Figs.~\ref{entangled}(b)-(c)]. Simulations conducted by integrating the Schr\"odinger equation for a three-spin system show excellent agreement with the data [Fig.~\ref{entangled}(d)], confirming that we have successfully transferred one member of the entangled pair to a distant electron. 

We note that such a $\Delta B$ measurement across two dots is a routine procedure for singlet-triplet qubits when the two electrons are separated to neighboring dots via tunneling~\cite{Petta2005}. Recently, $\Delta B$ oscillations between singlet pairs separated to distant dots via tunneling have also been observed~\cite{Fujita2017,Flentje2017,Nakajima2018}. In the present work, however, we use repeated coherent SWAP operations to move quantum spin states instead of electrons. 

As a check, we have performed the same experiment while omitting both SWAP operations. In this case, we also observe oscillations around the magnetic gradient, but with a different characteristic frequency, corresponding to $\Delta B_{34}$ [Figs.~\ref{entangled}(b)-(c)]. Figure~\ref{entangled}(e) shows the time evolution of $\Delta B_{24}$ and $\Delta B_{34}$ during the course of the experiment. Since both field gradients result from different random nuclear spin ensembles, we expect their time evolution to be different, as we observe [Fig.~\ref{entangled}(e)]. 

When we omit only the final SWAP operation, we observe small-amplitude oscillations, consistent with our simulations [Figs.~\ref{entangled}(c)-(d)]. A perfect initial $S_{23}$ operation would completely transfer the entanglement between dots 3 and 4 to dots 2 and 4, and we would not expect to observe oscillations without a final $S_{23}$. However, our $S_{23}$ operation is imperfect, because the magnetic gradient $\Delta B_{23}$ prevents a pure exchange rotation. After this imperfect SWAP operation, the electron spin in dot 3 remains weakly entangled with the electron spin in dot 4, and weak $\Delta B_{34}$ oscillations are observed. As expected, the oscillation frequency in this case clearly corresponds with $\Delta B_{34}$ [Fig.~\ref{entangled}(e)]. To ensure that differences between these three cases do not result from a randomly changing nuclear magnetic field between experiments, but instead result from the transmission of entangled states, we have interleaved the averaging of these measurements in time (see Methods). 

We can also coherently transfer one spin state of an entangled singlet pair to the other end of the array and back. We achieve this by applying $S_{23}$ and $S_{12}$ operations before and after free evolution around the magnetic gradient (Extended Data Figure~\ref{4dotEntangle}). 

The primary limiting factors of the spin-state transfer operation are the presence of a magnetic gradient between the dots and the temporal fluctuations in this gradient, resulting from the nuclear spin noise. In general, exchange coupling tends to swap the state of two spins, but a magnetic gradient $\Delta B$ tends to drive transitions to the singlet or unpolarized triplet configurations of the two spins~\cite{Petta2005}. The presence of a magnetic gradient therefore makes single-pulse pure exchange rotations impossible. In this work, we minimized this effect by using exchange strengths of several hundred MHz. Typical gradient strengths were several tens of MHz. In addition, because we do not perform pure exchange rotations, the final joint spin state of a pair of spins after a SWAP operation is also not an eigenstate of the local magnetic gradient. Thus, the magnetic gradient causes continued unwanted evolution of the joint spin state. Finally, the nuclear spin fluctuations can also create second-order noise in the exchange splitting. 

As described in Methods, we included these effects, in addition to charge noise, in numerical simulations of the coherent spin-state transfer, and we find good agreement with our data, as shown in Extended Data Figs.~\ref{3dotSim} and ~\ref{4dotSim}. Based on our simulations, we expect that the state fidelity after a SWAP operation for single-spin eigenstates is approximately 0.90 (see also ~Extended Data Fig. \ref{swapCal}). The error results almost entirely from the nuclear magnetic gradient. We estimate that the state fidelity after a SWAP operation on a singlet state is about 0.65. However, the infidelity in this case largely results from the gradient-induced evolution from the singlet state to unpolarized triplet during the SWAP. If the gradient is stable, a fidelity of about 0.9 can be recovered by adding a free evolution period after the SWAP pulse, as demonstrated in Extended Data Fig.~\ref{entangledFid} and suggested by Fig.~\ref{entangled}. 

The fidelity of spin-state transfer via Heisenberg exchange can be improved by minimizing the magnetic gradient. In particular, we expect that exchange-based spin-state transfer will work even better in silicon qubits, where nuclear spin fluctuations are suppressed. When magnetic gradients are needed, it is likely that resonant approaches and dynamically corrected exchange gates could be used to implement high-fidelity exchange rotations (see Methods).

We have demonstrated coherent spin-state transfer via Heisenberg exchange by transmitting the spin state of an electron back and forth along an array of electrons in a quadruple quantum dot. We have transfered single-spin eigenstates and entangled states via coherent SWAP gates between all neighboring pairs of spins in a four-qubit array. In the future, we expect that spin-state transfer via exchange will be useful in spin-based quantum computing for multi-qubit gates and quantum error correction in large spin-qubit arrays. Our work illustrates how it is possible to transmit the quantum state of an object without moving the object itself and provides a vivid example of the exciting and intriguing potential of quantum physics for the transmission, storage, and manipulation of information.

%

\section{Acknowledgments}
This research was funded by the the University of Rochester, ARO/LPS through Grant No. W911NF-17-1-0260, and the DARPA DRINQS program through Grant No. D18AC00025. 

\section{Author Contributions}
Y.P.K, H.Q, and J.M.N. fabricated the device and performed the experiments. S.F., G.C.G., and M.J.M. grew and characterized the AlGaAs/GaAs heterostructure. All authors discussed and analyzed the data and wrote the manuscript.

\section{Methods}
\subsection{Device}
The quadruple quantum dot is fabricated on a GaAs/AlGaAs hetereostructure with a two-dimensional electron gas located 91 nm below the surface. The two-dimensional electron gas density $n=1.5 \times 10^{11}$cm$^{-2}$ and mobility $\mu=2.5 \times 10^6$cm$^2/$Vs were measured at $T=4$K. Voltages applied to four layers of overlapping Al depletion gates~\cite{Angus2007,Zajac2015,Zajac2016} define the quadruple-dot potential. In addition to the three layers of aluminum gates shown in Fig.~\ref{apparatus}, we also found it necessary to deposit a grounded top gate over the device, likely to screen the effects of disorder in the two-dimensional electron gas, perhaps imposed by the 10-nm-thick aluminum oxide layer we deposited via atomic layer deposition. The quadruple dot is cooled in a dilution refrigerator to a base temperature of approximately 10 mK. An external magnetic field B=0.5 T is applied in the plane of the semiconductor surface perpendicular to the axis connecting the quantum dots. This orientation of the magnetic field ensures effective dynamic nuclear polarization~\cite{Nichol2015}.

We tune the device to the single occupancy regime, in which each dot is occupied by a single electron. The tune-up process is greatly facilitated by the use of ``virtual gates''~\cite{Baart2016,Mills2019}, which enables independent adjustment of the chemical potentials of the quantum dots [Fig.~\ref{apparatus}(c)]. In our approach, we correct for the capacitive coupling of all barrier and plunger gates to the chemical potential of each dot. Changing the tunnel barrier between a pair of dots involves changes to that barrier gate and also the application of compensation pulses to the plunger gates to the keep the chemical potentials of the dots fixed.

The detunings for each pair of dots are defined as follows. The detuning of the left side $\epsilon_L=\mu_2$, with $\mu_1=-\mu_2$. $\epsilon_L=0$ at the (0,2)-(1,1) crossing. The detuning of the right side $\epsilon_R=\mu_3$, with $\mu_4=-\mu_3$. $\epsilon_R=0$ at the (0,2)-(1,1) crossing.  The detuning of the middle pair $\epsilon_M=0$ when $\epsilon_L=10$mV and $\epsilon_R=6$mV. Changes to the detuning of the middle pair are such that $\Delta \epsilon_M=\Delta \mu_2$, with $\Delta \mu_3=-\Delta \mu_2$.

\subsection{Initialization and readout}
To initialize the array, we configure it as a pair of singlet-triplet qubits~\cite{Petta2005,Foletti2009}. We load two electrons in the singlet configuration in dots 1 and 4 each via electron exchange with the reservoirs~\cite{Foletti2009}. If we diabatically separate the electrons, they remain in the singlet state. We can also adiabatically separate the electrons into neighboring dots such that each dot has one electron. Upon adiabatic separation, the singlet states evolve into product states, with one electron spin-up and the other spin-down in each pair. The orientation of the spins is determined by the local magnetic field gradient. In the present case, the magnetic gradient results from the hyperfine interaction between the electron and Ga and As nuclear spins, each of which have nuclear spin $3/2$~\cite{Taylor2007}. For this work, we empirically observe that the magnetic gradient of dots 1-2 is metastable, and the gradient usually favors spin down in dot 1 and spin up in dot 2. We use dynamic nuclear polarization and feedback~\cite{Foletti2009,Bluhm2010,Shulman2014,Nichol2015} to set the magnetic gradient of dots 3-4, such that the ground state is spin down in dot 3 and spin up in dot 4. We have used a sequence of exchange oscillation measurements to verify that the ground state of the quadruple dot array, initialized in this way, is $\bra{\D \U \D \U}$, as we expect (Extended Data Fig.~\ref{statePrep}). It is also possible to initialize either pair of quantum dots in the $\left|T_+\right\rangle=\left|\uparrow \uparrow \right\rangle$ configuration by electron exchange with reservoirs~\cite{Foletti2009} when the ground state of a pair of electrons has one electron in each dot.

The assumption of metastability of the left-side gradient does not affect the data. It only affects our prediction for the measurement outcomes. If this assumption were violated at any time, it would appear to diminish the apparent agreement between our data and our predictions.

The prepared triplet states in Fig.~\ref{teleport} do not appear with perfect visibility. This reduction in visibility occurs because the Zeeman energy of the electrons at $B=0.5$T is $g \mu_B B/k_B\approx 100$ mK, which is not significantly higher than the thermal energy, and excited spin states remain populated to a small degree. Increasing the magnetic field or decreasing the temperature could improve the triplet visibility.

After manipulating the spins, we read them out by adiabatically moving the electrons in dots 1 and 2 both into dot 1, and we move the electrons in dots 3 and 4 into dot 4. If the joint spin state of each pair evolves into the singlet state during adiabatic transfer, both electrons can occupy the same dot~\cite{Petta2005}. However, if the pair evolves to a triplet state (if they have the same spin, for example), the Pauli exclusion principle forbids both electrons from occupying the ground state of the outer dot, and the pair remains separated. We detect this change in charge configuration through rf-reflectometry of proximal sensor quantum dots~\cite{Barthel2010}. In addition to conventional Pauli spin-blockade, we also use a shelving mechanism to increase the visibility of the readout~\cite{Studenikin2012}, and we can achieve single-shot readout within $5\mu$s integration times. 

All data presented here were taken by reading out the two sides of the array sequentially. Specifically, for each single shot measurement, we read out only one side. Although we applied exactly the same initialization and exchange pulse sequence when reading out different sides, we apply a different readout sequence depending on the side, as we discuss below. Sequential readout of the sides is sufficient to demonstrate transmission of single-spin eigenstates, because single-shot correlations are not required. 

We observe that reading out both sides of the array during the same single-shot measurement results in significant state-dependent crosstalk on the left-side signal from the right side. This effect results from the capacitance between the right and left sides of the array. Although the idling configuration of each side is in the (0,2) charge configuration, exchange pulses cause each side sometimes to occupy the (1,1) charge configuration due to Pauli spin blockade, and we do not know ahead of time which charge configuration each side will have for a given single-shot measurement. Changes in the charge configuration of one side shift the charge stability diagram of the other side, and these shifts interfere with the measurement process. In particular, we observe that when the right side is in the (1,1) charge configuration, the left side experiences rapid relaxation to the (0,2) singlet state during adiabatic transfer to the readout position. We believe this results from inadvertent electron exchange with the reservoirs on the left side when the right side occupies the (1,1) charge configuration. 

We solve this problem by adiabatically transferring the left-side electrons only when the right side occupies the (0,2) charge configuration. Specifically, we adiabatically transfer from (0,2) to (1,1) on the left side before the right side, and we transfer back to (0,2) on the left side after the right side. 

In addition, to readout the left side, we reload the right side as an (0,2) singlet after adiabatic transfer to the right-side readout position but before adiabatic transfer to the left-side readout position. This step ensures that the right side has the same charge configuration every time the electrons on the left side separate and recombine. We verified that the presence or absence of this reload on the right side has no discernible effect on left-side exchange measurements, and it removes the state-dependent crosstalk effect. To readout the right side, we omit the extra reload and enforce a wait for the same length of time. We use this protocol to take the data for all panels in the paper. We emphasize that the extra initialization step on the right side always took place after all exchange pulses were finished, and exactly the same initialization and exchange pulses were applied in the sequences to readout both the right and left sides of the array.

We have also observed similar crosstalk effects from the left side on the right side. In general, we observe that crosstalk effects depend sensitively on device tuning and may also partly result from an imperfect gate capacitance matrix. However, for the tuning used for the experiments described here, left-to-right crosstalk was not significant. 

To demonstrate transfer of entangled states, we only measured the right side of the array. Because we measure the right side directly in the singlet-triplet basis, measurement of a single side is sufficient to distinguish evolution between these entangled states. 

\subsection{Exchange gates}
We induce exchange coupling between two electrons by applying a voltage pulse to the barrier gate between them~\cite{Reed2016,Martins2016}. The voltage pulse creates an overlap between the wavefunctions of neighboring electrons, which causes them to evolve under exchange. Barrier-induced exchange coupling between fully separated electrons is first-order insensitive to charge noise~\cite{Martins2016,Reed2016} associated with the plunger gates, which would otherwise randomly shift the locations of the electronic wavefunctions and promote rapid decoherence. This insensitivity to noise is critical for the results we describe in the main text. We have empirically found that the overlapping gate architecture is essential for high-fidelity barrier-controlled exchange gates. During the barrier pulses, we apply compensation pulses to the plunger gates to keep the chemical potentials of the dots fixed~\cite{Martins2016}. 

\subsection{Interleaved measurements}
We interleaved the averaging of different pulse sequences to demonstrate transmission of entangled states. The purpose of interleaving the measurements was to ensure that changing nuclear fields did not confound the measurement, since we rely on observing coherent oscillations of different frequencies. Specifically, we performed 32 single shot experiments (initialization, evolution, and measurement), each lasting 28 $\mu$s, omitting both $S_{23}$ operations. Immediately following this set, we performed 32 single-shot measurements omitting only the second $S_{23}$ operation, and then we performed 32 single-shot measurements with both $S_{23}$ operations. We then averaged each set of 96 measurements 512 times, and the averaged result is displayed as one line in Fig.~\ref{entangled}(b). Each line takes approximately 1 second. Empirically, we find nuclear magnetic fields are reasonably stable on this timescale~\cite{Shulman2014}. We repeated this process 16 times. As can be seen in Fig.~\ref{entangled}(b), each line shows coherent $\Delta B$ oscillations. It is also evident that averaging all lines together would show significant dephasing. Note that in this experiment, we stabilized $\Delta B_{34}$ using nuclear pumping. 

\subsection{Simulation}
To generate the simulation in Fig.~\ref{entangled}(d), we numerically integrated the Schr\"odinger equation for a three spin system. We generated a simulated SWAP operation from the following Hamiltonian: 

\begin{align}
	H&=\frac{h}{4}J_{23} ( \sigma_{x,2} \otimes \sigma_{x,3}+\sigma_{y,2} \otimes \sigma_{y,3} + \sigma_{z,2} \otimes \sigma_{z,3})\\ \nonumber
	&+\frac{g \mu_B}{2}\sum_{k=2}^4 B_k \sigma_{z,k} \label{Hamiltonian}
\end{align}

We assumed a fixed exchange coupling of $J_{23}$ of 150 MHz between spins $2$ and $3$, and we adjusted the time for the SWAP operation to give a $3\pi$ pulse. These parameters correspond closely to the actual experiments. We adjusted the local nuclear magnetic fields $B_k$ of spin $k$ to be $(0,75,35)$ MHz $\times \frac{2 h}{g \mu_B}$ in dots 2-4. These were adjusted to match the frequencies observed in Fig.~\ref{entangled}(c). 

We initialized the three-spin system in the $|\uparrow\rangle \otimes \frac{1}{\sqrt{2}} \left( |\uparrow \downarrow\rangle - |\downarrow \uparrow \rangle \right)$ state, corresponding to dots 2-4. After applying an $S_{23}$ operation (including the effects of magnetic fields), we evolved the system for a variable evolution time in the presence of the magnetic fields. Then we applied a final $S_{23}$ operation, and we projected the resulting state along all states with a singlet in dots 3 and 4. To generate the simulated control measurements, we omitted the relevant $S_{23}$ operations.

We have also simulated the single-spin transfer in a similar way. We numerically integrated the Schr\"odinger equation for a 4-spin system. We choose the nuclear magnetic fields for each site to be approximately $(0,50,100,150)$ MHz $\times \frac{2h}{g \mu_B}$, for the three-dot transfer and $(0,30,60,90)$ MHz $\times \frac{2h}{g \mu_B}$ for the four-dot transfer. Choosing different gradient configurations more accurately reproduces the experimental results. Both gradient configurations fall within the expected range of natural gradient fluctuations. We allow the gradient to fluctuate by 30 MHz on each dot between runs. This is slightly larger than the expected ~20 MHz gradient fluctuations~\cite{Nichol2015}, perhaps due to unintentional nuclear polarization. We assume $J=200$ MHz for each exchange pulse, and we set the pulse time to generate a 3$\pi$ rotation. We also include a 3-10 ns wait between exchange pulses, which we used in the experiments. We also included thermal population of excited states by during the $T_+$ loading process by assuming an electron temperature of 100 mK~\cite{Orona2018}. We applied the pulse sequences described in the main text, and then we projected the left and right sides onto final states with the $\bra{\D \U}$ configuration on either side, corresponding to the singlet outcome after adiabatic charge transfer. We averaged the simulation results over approximately 100 realizations of the noise. The results of this simulation are shown in Extended Data Figs.~\ref{3dotSim} and ~\ref{4dotSim}. 

\subsection{Fidelity estimate}
We estimate the fidelity of the SWAP operation for single-spin eigenstates by simulating the effect of a realistic $S_{23}$ operation on an initial state $\bra{\psi_0}=\bra{\U \U \D \U}$. We simulate the $S_{23}$ operation as described above. The ideal target state after this operation is $\bra{\psi_t}=\bra{\U \D \U \U}$. We compute an estimated state fidelity $F=\left|\left\langle \psi_t|S_{23}|\psi_0 \right \rangle \right|^2$, where $S_{23}$ is generated by exponentiating a 4-spin Hamiltonian, obtained by extending the sum in Eq.~\ref{Hamiltonian} to all 4 spins. We averaged the resulting fidelity over 2000 different realizations of magnetic and electrical noise. Based on observed exchange quality factors, we included quasi-static fractional electrical noise $\delta_{J_{23}}/J_{23}=0.02$ in the simulation. Including these effects, we calculate $F\approx 0.90$. Eliminating the magnetic fluctuations while retaining the static gradient increases the fidelity to approximately 0.94, and eliminating both the magnetic fluctuations and the static gradient (leaving only charge noise) improves the fidelity to above 0.99. The sensitivity to magnetic noise decreases as the static gradient decreases because the overall probability that the magnetic gradient will approach the exchange coupling diminishes in this case.

To assess the fidelity of the SWAP operation for entangled states, we begin with the initial state $\bra{\psi_0}=\frac{1}{\sqrt{2}}\left(\bra{\U \U \U \D}-\bra{\U\U\D\U}\right)$. After a perfect $S_{23}$ operation, the target final state is $\bra{\psi_t}=\frac{1}{\sqrt{2}}\left( \bra{\U\U\U \D}-\bra{\U\D\U\U}\right)$. We calculate and average the fidelity as described above and also find $F\approx0.65$. In this case, the infidelity largely results from coherent evolution of the singlet state to the unpolarized triplet state. This evolution can be undone by adding allowing the state to evolve under the action of the magnetic gradient following the $S_{23}$ operation. If the gradient is static, the triplet will return to the singlet state after some time. The return of the singlet is evident in Fig.\ref{entangled}(b)-(c). We have also explicitly simulated the effects of adding a free-evolution period in Extended Data Fig.~\ref{entangledFid}. We find that the singlet fidelity reaches a maximum of about 0.9. 

In the future, spin state transfer via Heisenberg exchange will work best in systems with small gradients and small levels of spin noise, such as silicon qubits. However, dynamically corrected gates ~\cite{Wang2012} and resonant approaches~\cite{Nichol2017,Sigillito2019} can also be used to implement high-fidelity SWAP operations in the presence of gradients and noise.

\subsection{Data Availability}
The data that support the findings of this study are available from the corresponding author upon reasonable request.

\renewcommand{\figurename}{Extended Data Figure}
\renewcommand{\thefigure}{\arabic{figure}}
\setcounter{figure}{0}    

\newpage
\begin{figure*}
	\centering
	\includegraphics{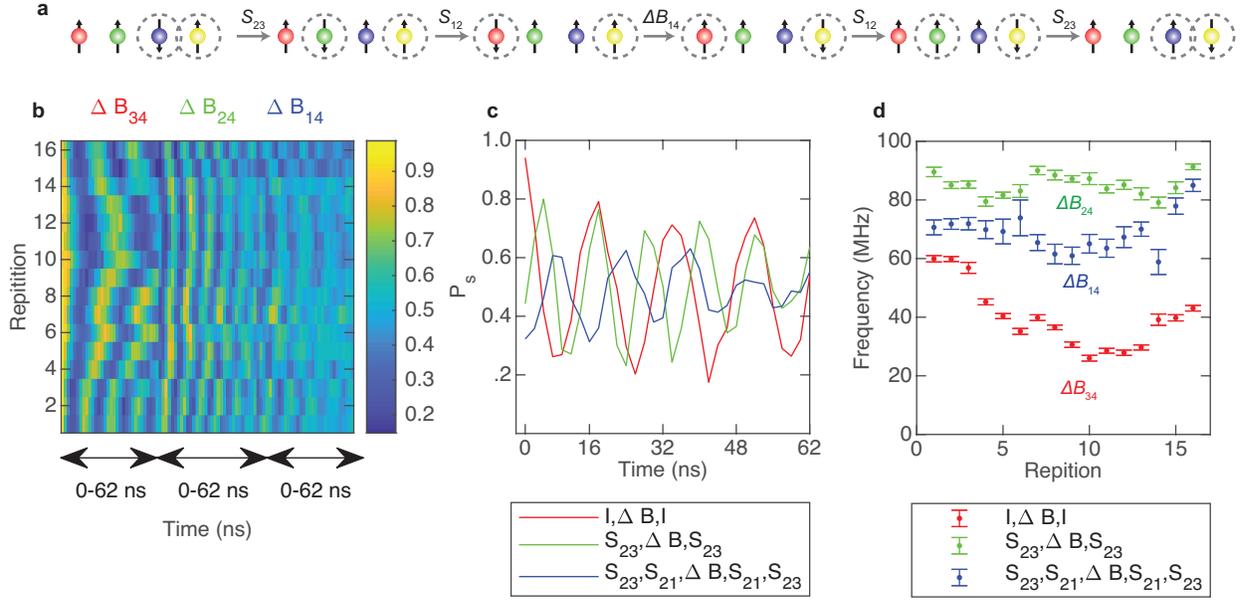}
	\caption{\label{4dotEntangle} Experimental data showing four-dot transfer of entangled states. (a) Schematic of the four-dot entangled state transfer process. (b) Interleaved data showing $I,\Delta B, I$; $S_{23},\Delta B, S_{23}$; $S_{23},S_{12},\Delta B,S_{12}, S_{23}$ measurements. (b) Data from repetition 2, plotted on the same horizontal axis. (c) Time evolution of the different magnetic gradients. Because they result from different nuclear spin configurations, they have different values and time evolutions. Error bars are fit errors.}
\end{figure*}
\newpage

\begin{figure*}
	\centering
	\includegraphics{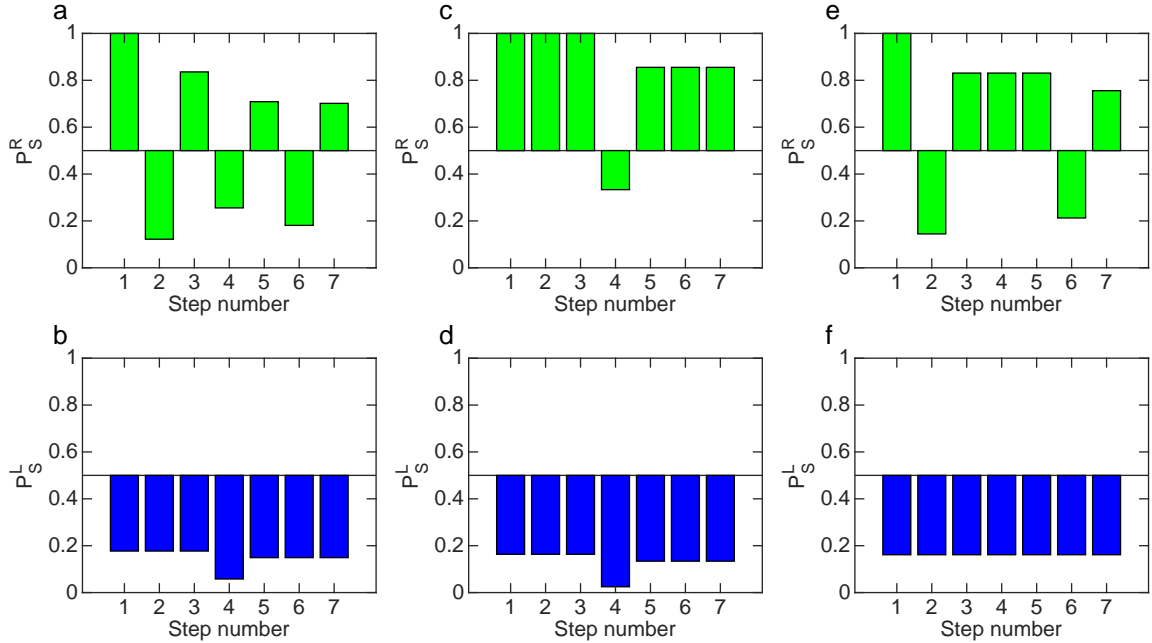}
	\caption{\label{3dotSim}  Results of the three-dot state-transfer simulation, as described in the Methods section, showing good agreement with the data in Fig.~\ref{teleport}. (a) Simulated right-side measurements for the $S_{34}, S_{34}, S_{23}, S_{23}, S_{34}, S_{34}$ sequence. (b) Simulated left-side measurements for the same sequence. (c) Simulated right-side measurements for the three-dot state transfer control sequence with $I$ in place of $S_{34}$. (d)  Simulated left-side measurements for the same sequence. (e) Simulated right-side measurements for the three-dot control sequence with $I$ in place of $S_{23}$. (d) Simulated left-side measurements for the same sequence. }
\end{figure*}
\newpage

\begin{figure*}
	\centering
	\includegraphics{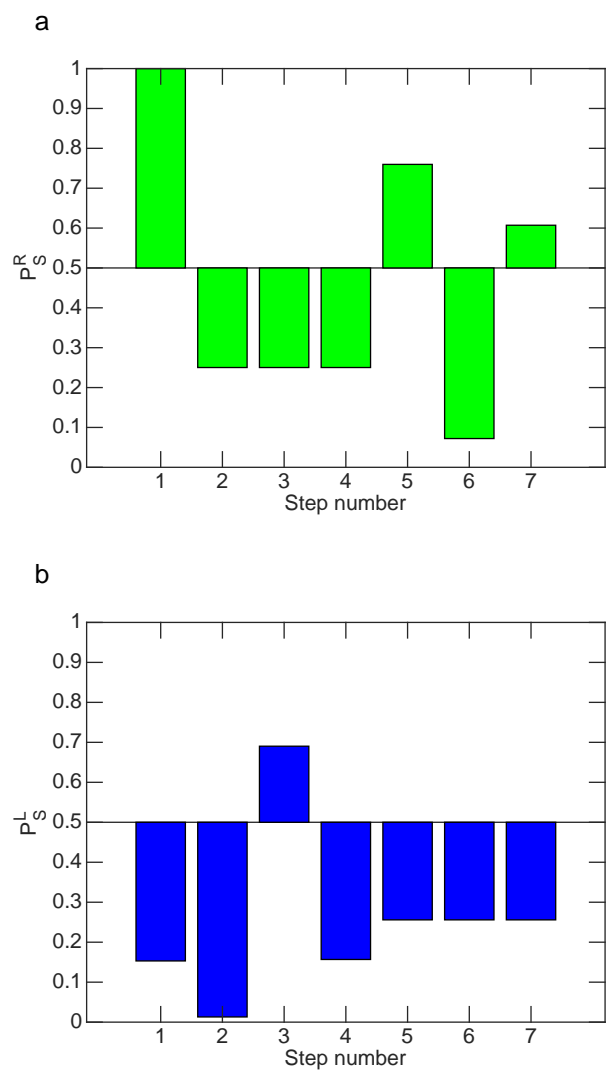}
	\caption{\label{4dotSim} Results of the four-dot state-transfer simulation, as described in the Methods section, showing good agreement with the data in Fig.~\ref{teleport}. (a) Simulated right-side measurements for the $S_{23}, S_{12}, S_{12} , S_{23}, S_{34}, S_{34}$ sequence. (b) Simulated left-side measurements for the same sequence. }
\end{figure*}
\pagebreak

\begin{figure*}
	\centering
	\includegraphics{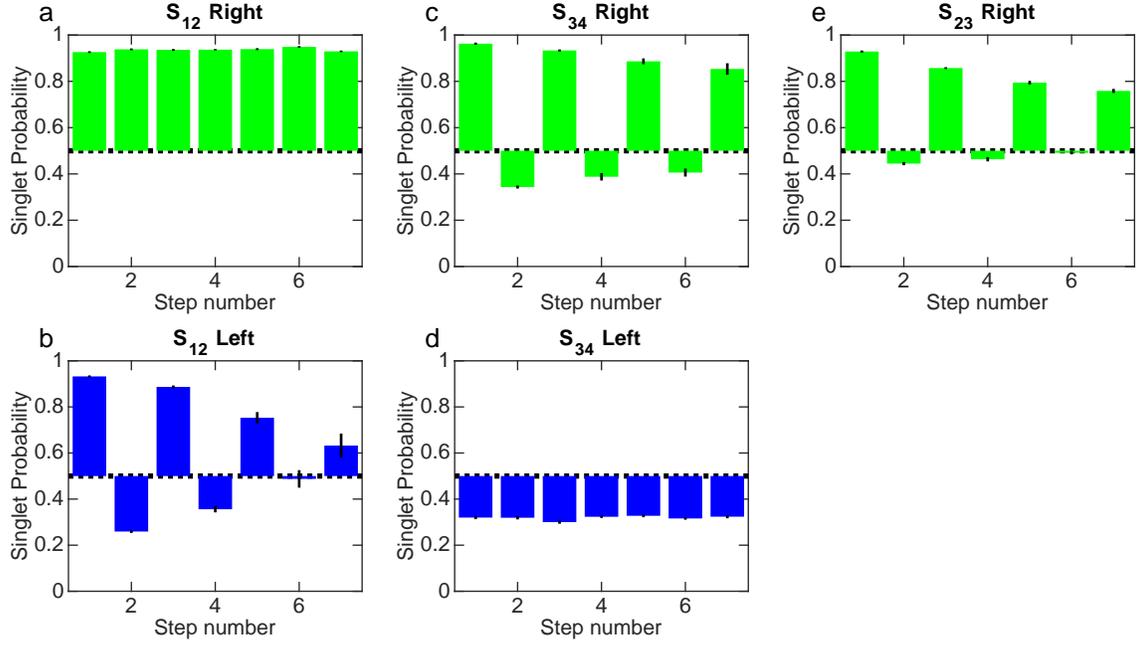}
	\caption{\label{swapCal} Calibration of SWAP operations by pulse concatenation. Each panel shows the results of concatanating specific operations. Each SWAP operation is implemented by a separate voltage pulse to a barrier gate. (a) Right-side measurements for repeated $S_{12}$ operations. Prior to the first step, the array was initialized in the $\bra{\D \U \D \U}$ state. (b) left-side measurements for repeated $S_{12}$ operations. (c) Right-side measurements for repeated $S_{34}$ operations. Prior to the first step, the array was initialized in the $\bra{\D \U \D \U}$ state. (d) Left-side measurements for repeated $S_{34}$ operations. (e) Right-side measurements for repeated $S_{23}$ operations. The array was initialized in the $\bra{\U \U \D \U}$ state. We did not record left-side measurements for this sequence. In all panels, vertical black lines indicate error bars. The error bars represent the standard deviation of 64 repetitions of the average of 64 single-shot measurements of each pulse configuration.}
\end{figure*}

\begin{figure*}
	\centering
	\includegraphics{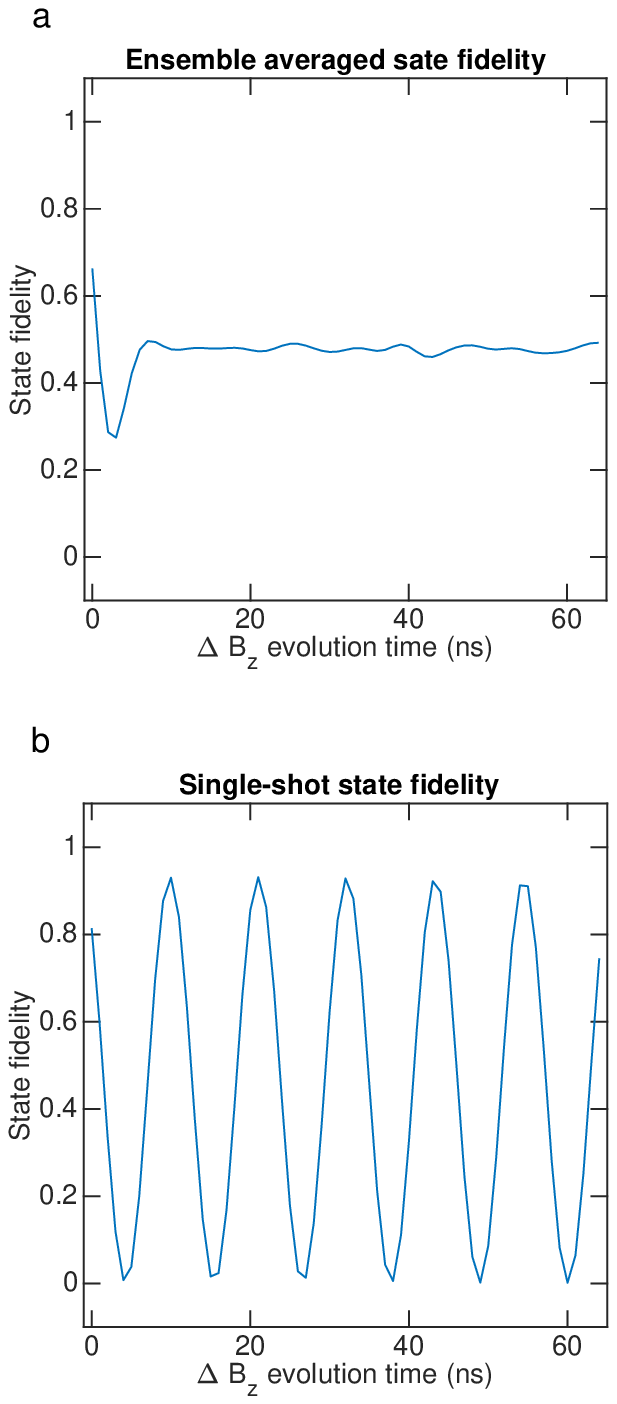}
	\caption{\label{entangledFid} Simulated fidelity of SWAP pulses for entangled states. (a) Simulated ensemble averaged state fidelity after applying a simulated realistic $S_{23}$ operation to the initial state $\bra{\psi_0}=\frac{1}{\sqrt{2}}\left(\bra{\U \U \U \D}-\bra{\U\U\D\U}\right)$. The target state is $\bra{\psi_t}=\frac{1}{\sqrt{2}}\left( \bra{\U\U\U \D}-\bra{\U\D\U\U}\right)$. The x-axis represents the free-evolution time of the state under the influence of the magnetic gradient after the exchange operation. The fidelity is averaged over 2000 different realizations of magnetic and electrical noise. The state fidelity has a maximum of about 0.65, and it quickly decays to 0.5. The decay results from the fluctuating magnetic gradient. (b) Simulated characteristic single-shot simulated state-fidelity for one realization of the noise. For specific times, the state fidelity returns to about 0.9. The magnetic gradient is assumed to be stable in each realization of the sequence. }
\end{figure*}

\begin{figure*}
	\centering
	\includegraphics{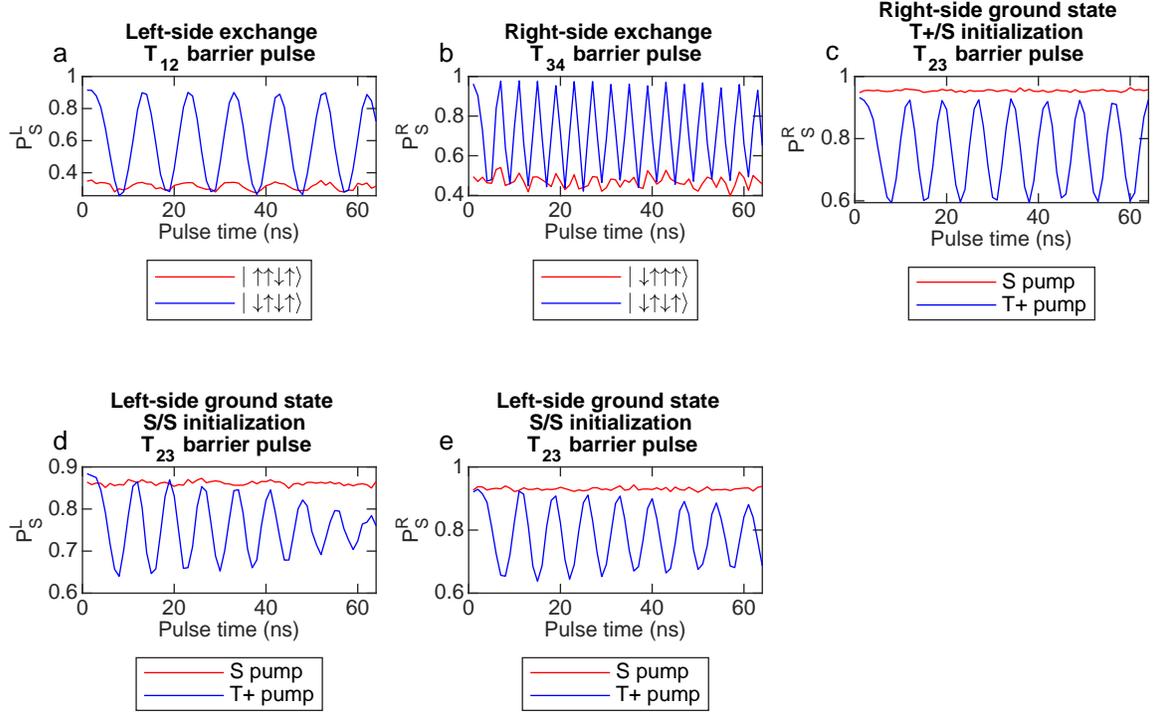}
	\caption{\label{statePrep} Quadruple quantum dot state preparation. (a) Verification of exchange oscillations on the left side. Initializing the left side in the $\bra{\U \U}$ state before a $T_{12}$ pulse yields no exchange oscillations. Initialization in the $\bra{\D \U}$ state shows exchange oscillations. (b) Initializing the right side in the $| \uparrow \uparrow\rangle$ state before a $T_{34}$ pulse yields no exchange oscillations. Initialization in the $| \downarrow \uparrow\rangle$ state shows exchange oscillations. (c) Proof of the ground state orientation of the right side. We load the left side in the $| \uparrow \uparrow \rangle$ state and the right side by adiabatic separation of the singlet state, which gives either $\bra{\U \D}$ or $\bra{\D \U}$, depending on the sign of the gradient. We pulse $T_{23}$ to induce exchange between the middle two spins. Dynamic nuclear polarization with singlets yields no oscillations, while pumping with triplets yields oscillations. These data confirm that the separated singlet state evolves to the $\bra{\D \U}$ state under triplet pumping for the right side. (d) Proof of the ground state of the left side. We initialize the array by separating singlets on both sides. In the case of triplet pumping on the right side, the third spin is $\bra{\D}$, so it must be the case that the second spin is $\bra{\U}$ in order to generate exchange oscillations with a $T_{23}$ pulse, as measured on the left side. Singlet pumping on the left side yields no exchange oscillations. (e) The same initialization and pulses as (e), but measured on the right side. In all cases, $P_S^{L(R)}$ indicates the singlet return probability measured on the left(right) side.}
\end{figure*}
\newpage

\end{document}